\documentclass[12pt]{article}
\usepackage{hyperref}
\usepackage{graphicx,amsmath,amssymb}

\newcommand\pubnumber{JLAB-PHY-18-2855}
\newcommand\pubdate{\today}

\def\JLab{Thomas Jefferson National Accelerator Facility \\ Newport News, VA 23606, USA}
\def\support{\footnote{This work is based in part upon work supported by the U.S. Department of Energy, 
the Office of Science, and the Office of Nuclear Physics under contract DE-AC05-06OR23177.}}

\def\Title#1{\begin{center} {\Large #1 } \end{center}}
\def\Author#1{\begin{center}{ \sc #1} \end{center}}
\def\Address#1{\begin{center}{ \it #1} \end{center}}

\newcommand\pubblock{\rightline{\begin{tabular}{l} \pubnumber\\
         \pubdate  \end{tabular}}}
\newenvironment{Abstract}{\begin{quotation}  }{\end{quotation}}
\newenvironment{Presented}{\begin{quotation} \begin{center} 
             PRESENTED AT\end{center}\bigskip 
      \begin{center}\begin{large}}{\end{large}\end{center} \end{quotation}}

\def\beq{\begin{equation}}
\def\eeq#1{\label{#1}\end{equation}}
\def\eeqn{\end{equation}}

\def\beqa{\begin{eqnarray}}
\def\eeqa#1{\label{#1}\end{eqnarray}}
\def\eeqan{\end{eqnarray}}

\let\bar=\overbar

\def\Dslash{\not{\hbox{\kern-4pt $D$}}}
\def\dslash{\not{\hbox{\kern-2pt $\del$}}}

\def\msb{{\bar{\ssstyle M \kern -1pt S}}}

\begin{document}
\begin{titlepage}
\pubblock

\vfill
\Title{Nucleon spin structure measurements at Jefferson Lab}
\vfill
\Author{ Alexandre Deur\support}
\Address{\JLab}
\vfill
\begin{Abstract}

We summarize the measurements investigating the nucleon spin structure  done at Jefferson Lab, a multi-GeV 
continuous electron beam facility located in Newport News, Virgina, USA. 
After motivating spin structure studies, we explain how Jefferson Lab uniquely contributes to them and describe
the experimental program that was achieved with energies up to 6 GeV.
We then discuss the continuation of this program at the higher energies now available
thanks to the recent 12 GeV upgrade of Jefferson Lab's accelerator.
\end{Abstract}
\vfill
\begin{Presented}
Thirteenth Conference on the Intersections of Particle and Nuclear Physics\\
Palm Spring, CA,  USA, 05/29-06/03 2018
\end{Presented}
\vfill
\end{titlepage}
\def\thefootnote{\fnsymbol{footnote}}
\setcounter{footnote}{0}

\section{Introduction: why study the nucleon spin ?}

Spin degrees of freedom (d.o.f) in hadron structure studies have been 
actively  used for more than 40 years~\cite{Deur:2018roz}.
Additional d.o.f test comprehensively a
theory, as famously expressed by Bjorken~\cite{Bjorken:1996dc}:
``Polarization data have often been the graveyard of fashionable theories. If theorists had their way,
they might well ban such measurements altogether out of self-protection." 
In fact, Bjorken's sum rule~\cite{Bjorken:1966jh} played the leading role in establishing
that perturbative QCD (pQCD) correctly describes the strong force at high energy, even
when spin d.o.f are explicit. 

Beside validating pQCD, one also needs to understand QCD at low energy, i.e. in its nonperturbative
region. One approach is to study the origin of the nucleon spin, i.e. what  
values have the quantities in the right side of the nucleon spin sum rule 
\vspace{-0.3cm}
\begin{equation}
\vspace{-0.3cm}
J = \frac{1}{2}= \frac{1}{2}\Delta\Sigma+L_q + \Delta G + L_g,
\label{eq:spin SR}
\end{equation}
as they all have a  nonperturbative origines. Here, $J$ is the nucleon spin, 
$\Delta\Sigma/2$ is the contribution from the quark spins,
$L_q$ the contribution from their orbital angular momenta (OAM),  $\Delta$G comes from 
the gluon spins and $L_g$ from the gluon OAM. 
The success of the nonrelativistic constituent quark model in 1970s-1980s to describe the unpolarized nucleon 
structure suggested that only the quark spins were relevant, i.e. $J \approx \Delta\Sigma/2$. 
However, a first measurement yielded $\Delta\Sigma \approx 0$~\cite{Ashman:1987hv} 
showing that the nucleon spin composition is not as trivial as what the constituent quark 
model suggested. This complexity signals that interesting information on the nucleon 
structure and on the  strong force nonperturbative aspects are revealed by spin studies.                                                                       
Particularly, information can be gained on quark  confinement and the emergence of effective d.o.f 
(hadrons) from fundamental ones (partons), as discussed
in Section~\ref{AdS/QCD}. Such gains come from comparing data to nonperturbative
calculations obtained using lattice calculations, the Dyson-Schwinger equations~\cite{the:SDE review}, 
or effective approaches like chiral perturbation
theory ($\chi$PT)~\cite{Bernard:2006gx} or gauge-string duality (AdS/QCD)~\cite{Brodsky:2014yha}.

Another benefit of nucleon spin studies is that accurate parton distribution functions 
(PDFs) are needed for high energy research  (e.g. for
Standard Model tests and searches of new physics) or, on the other side of the
energy scale, for atomic physics.

%We will see that JLab is contributing to all these aspects of the study on the nucleon spin structure.

\section{The 6 GeV Jefferson Lab nucleon spin program}

We will mostly discuss here the inclusive Jefferson Lab (JLab) 
program. For reports on polarized semi-inclusive deep inelastic scattering (SIDIS) at HERMES and on RHIC-Spin, 
see the proceeding contributions of B. Seitz, M. Liu and O. Eyser.

\begin{figure}[htb]
\centering
\vspace{-0.5cm}
\includegraphics[height=3.3in]{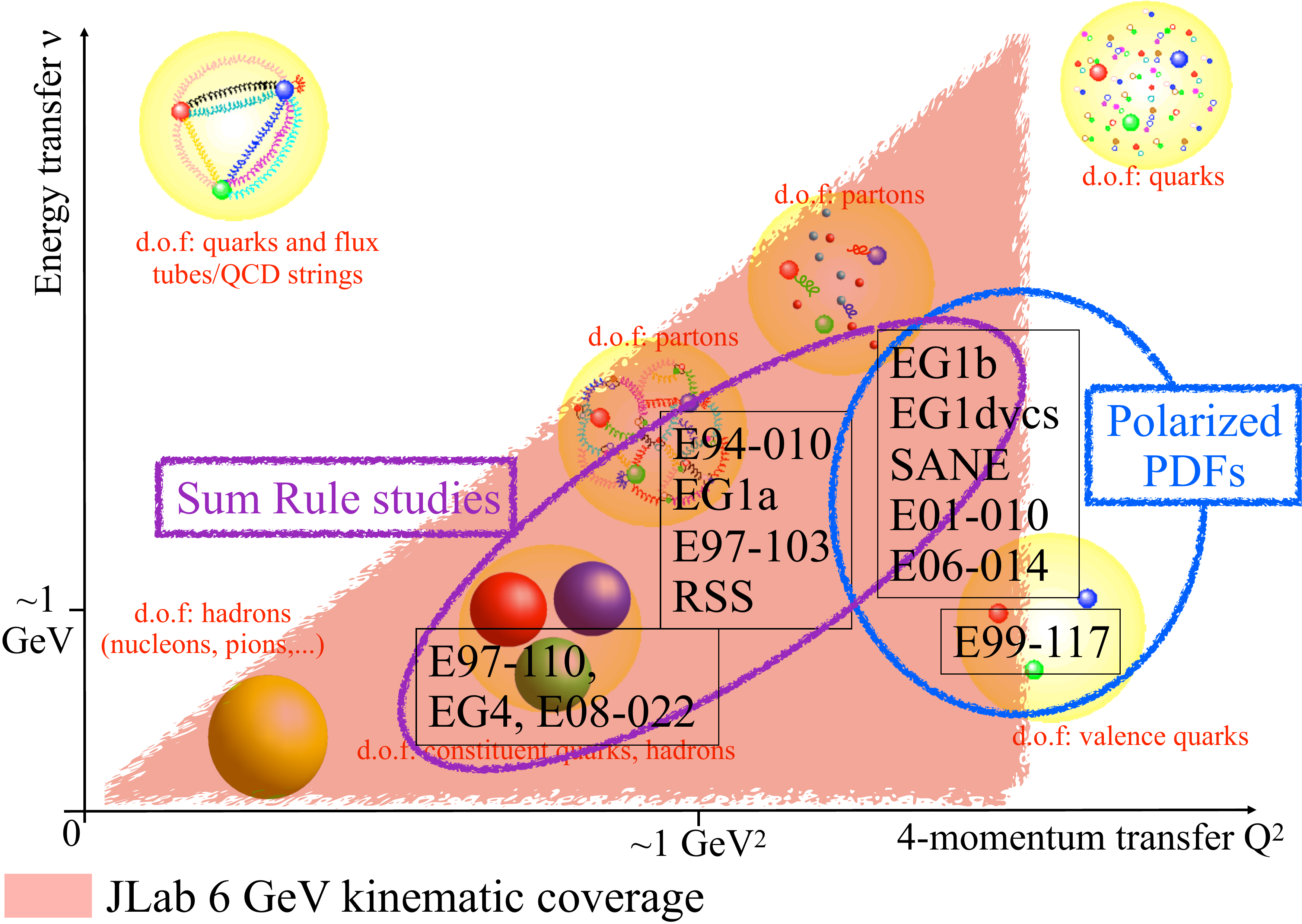}
\vspace{-0.45cm}
\caption{\small{Sketch of the $(Q^2,\nu)$ coverage of JLab at 6 GeV (pink area). The nucleon's changing aspects 
and relevant degrees of freedom are also pictured. In the squares are
the names of the inclusive polarized experiments. The ellipses indicate their main
topics.}}
\label{fig:kin6GeV}
\end{figure}

The kinematics of JLab at 6 GeV are sketched in Fig.~\ref{fig:kin6GeV} in the
energy transfer $\nu$ and momentum transfer squared $Q^2$ space. 
The relevant d.o.f, which depend on kinematics, are also pictured. 
In the Bjorken limit $\nu \to \infty$ and $Q^2 \to \infty$, with $x=Q^2/(2M\nu)$ finite 
($M$ is the nucleon mass), quarks are  free and are thus the relevant d.o.f.
At high but finite $\nu$ and $Q^2$ (deep inelastic scattering, DIS),  
gluons appear and also become  relevant d.o.f. As  $\nu$ and $Q^2$ decrease, 
gluons increasingly bind the quarks until those react coherently to the probing process.
Thus hadronic d.o.f arise, replacing the partonic ones.
\vspace{-0.3cm}

Also shown in Fig.~\ref{fig:kin6GeV}  are JLab's polarized inclusive  experiments and
the main topics they addressed: E94-010~\cite{Amarian:2002ar} and 
EG1~\cite{Yun:2002td}  chiefly studied sum rules in the parton to hadron transition region.
They were continued at lower $Q^2$ ($\chi$PT region) % ($Q^2 \ll M^2$) 
by  E97-110, E08-027, and EG4~\cite{Adhikari:2017wox}.
E97-103~\cite{Kramer:2005qe} and E06-014~\cite{Posik:2014usi} 
studied higher twists, i.e. the nonperturbative  $1/Q^{2n}$ corrections
arising once gluons link the PDF of the
struck quark to the PDF of the other partons, which are thus no more spectators to the 
probing process. RSS~\cite{Wesselmann:2006mw}, EG1dvcs~\cite{Prok:2014ltt}, 
E01-012~\cite{Solvignon:2013yun} and SANE~\cite{Armstrong:2018xgk}   measured 
the spin structure functions $g_1$ and $g_2$ in the DIS and resonance  regions. Finally,
E99-117~\cite{Zheng:2003un} investigated the high-$x$  region of DIS. Those are the
main goals, but high quality ancillary data were also typically provided, e.g. higher twist 
evaluations from E94-010 and EG1, or high precision high-$x$ DIS data from E06-014.

\subsection{PDF measurements}
Valence quarks dominate in the DIS high-$x$  region, which simplifies
the description of the nucleon structure. However, unpolarized PDFs  
decrease as $x \to 1$, which implies small cross-sections that are further suppressed by kinematic factors
varying at first order as $1/x$. The high polarized luminosity 
of JLab allowed to measure precise spin asymmetries in this region for the first time.
The relative simplicity of the high-$x$  region allows for non-trivial pQCD constraints on
the PDFs~\cite{Brodsky:1994kg}. Several model  predictions also exist,  including
constituent quark models~\cite{Isgur:1998yb},
the statistical model~\cite{Bourrely:2001du}, 
hadron-parton duality~\cite{Close:2003wz},
a Dyson-Schwinger approach~\cite{Roberts:2011wy}, 
bag models~\cite{Boros:1999tb}, and 
chiral soliton, instanton or covariant quark-diquark models~\cite{Kochelev:1997ux}.
 
\begin{figure}
%\vspace{-0.3cm}
\center
\includegraphics[scale=0.43]{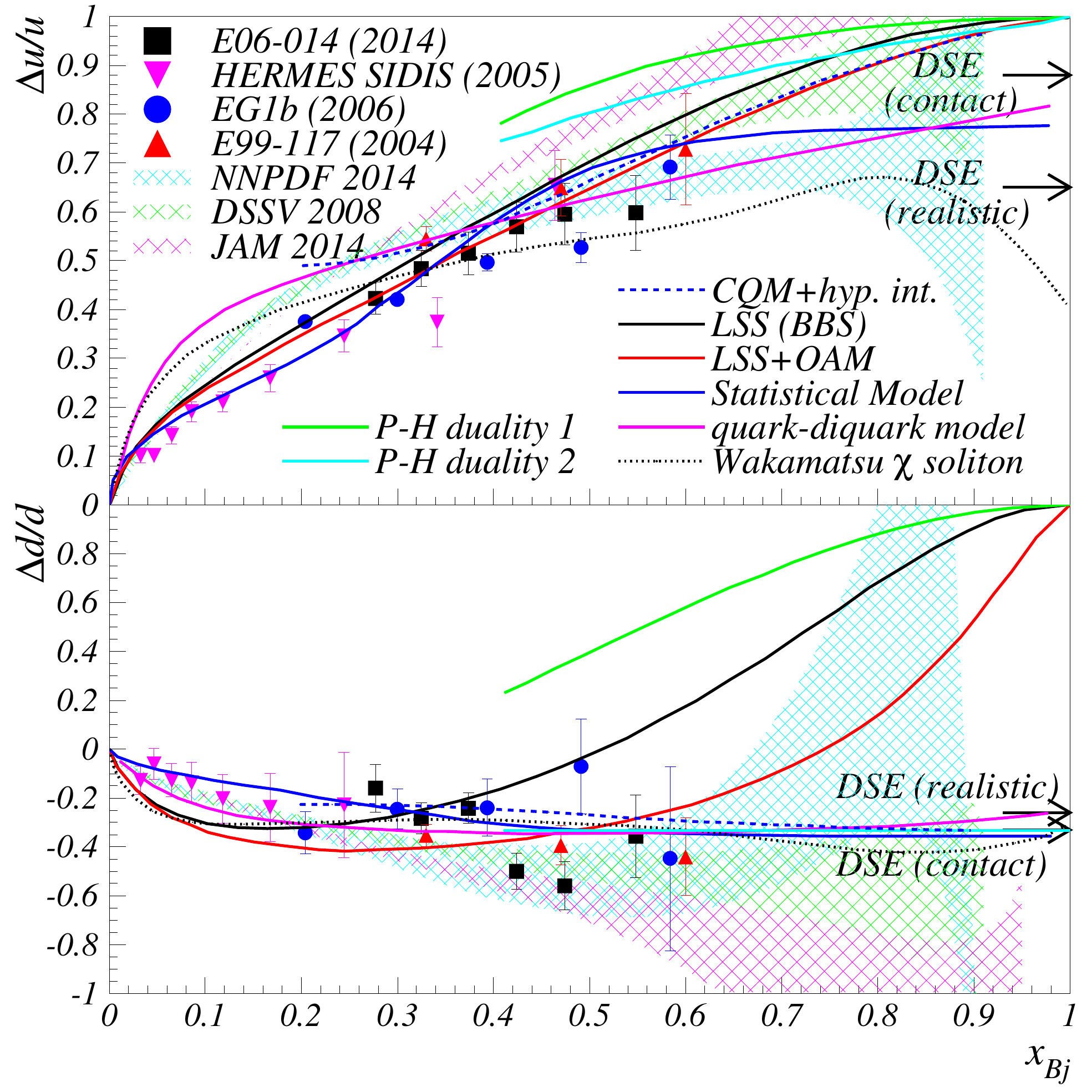}
\vspace{-0.6cm}
\caption{\label{fig:partons_polar_vs_x} \small{Data, global fits and model predictions for the quark polarizations
$\Delta q/q$. 
%The high polarized luminosity of JLab has enabled the precise study of nucleon spin structure at high-$x$.
}}
\end{figure}

Assuming  isospin symmetry and no sea quark, the quark polarizations $\Delta u/u$ and $\Delta d/d$ can be 
calculated from the high-$x$ spin asymmetry measurements: 
\vspace{-0.3cm}
 \begin{equation}
 \vspace{-0.2cm}
\frac{g_1^p}{F_1^p}=\frac{4\Delta u+\Delta d}{4u+d},~~~~~~~\frac{g_1^n}{F_1^n}=\frac{\Delta u+4\Delta d}{u+4d},
\end{equation}
where $F_1$ is the first unpolarized structure function. The quark polarizations are shown in 
Fig.~\ref{fig:partons_polar_vs_x} along with the predictions just 
listed and the NNPDF~\cite{Nocera:2014gqa}, 
DSSV~\cite{deFlorian:2009vb} and JAM~\cite{Jimenez-Delgado:2014xza} global fits.
Only the most precise high-$x$ data~\cite{Yun:2002td, Posik:2014usi, Zheng:2003un} are shown. 
At lower $x$ where sea quarks are significant, SIDIS reactions
are needed to separate quark flavors. The low $x$ data in Fig.~\ref{fig:partons_polar_vs_x} 
are from HERMES~\cite{Ackerstaff:1999ey}. 
A notable result at high $x$  is that $\Delta d/d<0$, 
contradicting pQCD unless OAM is present. 
This suggests that  $L_q$ in Eq.~(\ref{eq:spin SR}) is significant.

\subsection{Sum rules studies}

Another JLab contribution to nucleon spin studies is \emph{via} sum rules.
Those usually relate moments of structure functions %, a photoabsorption cross-section or a form-factor, 
to global properties of the studied particle. 
%This traditional definition has been generalized to also include
%double virtual Compton scattering amplitudes in addition to global properties. 
The most  famous spin sum rule is Bjorken's one~\cite{Bjorken:1966jh}. It
links the isovector part of $g_1$ to the nucleon axial charge $g_a$:
\vspace{-0.3cm}
 \begin{equation}
 \vspace{-0.2cm}
\Gamma_1^{p-n}(Q^2)\equiv \int_0^{1^-} g_1^p -g_1^n dx=\frac{g_a}{6}\bigg[1-\frac{\alpha_s}{\pi}-3.58\left(\frac{\alpha_s}{\pi}\right)^2
%-20.21\left(\frac{\alpha_{{\rm {s}}}}{\pi}\right)^{3} -175.7\left(\frac{\alpha_{{\rm {s}}}}{\pi}\right)^{4}
-O(\alpha_s) \bigg]+O(1/Q^2),
\label{eqn:bj}
\end{equation}
where $\alpha_s$ is QCD's coupling. The $\alpha_s$-terms are pQCD radiative 
effects~\cite{Kataev:1994gd}. They vanish for $Q^2 \to \infty$ where the sum rule was originally 
derived. The $O(1/Q^2)$ terms are higher twists, arising when $Q$ gets close
to the confinement scale $\Lambda_s$. The Bjorken sum rule derivation predates QCD
and thus is not a QCD prediction. However, its $\alpha_s$-dependence is a pQCD
prediction and the sum rule test at high $Q^2$ %, as done at SLAC, CERN and DESY,
verified that QCD is correct even when spin d.o.f are explicit.  

Another important spin sum rule is that of Gerasimov-Drell-Hearn (GDH)~\cite{Gerasimov:1965et}:
\vspace{-0.2cm}
\begin{equation}
\vspace{-0.2cm}
\int_{\nu_0}^{\infty}\frac{\sigma_{P}(\nu)-\sigma_{A}(\nu)}{\nu}d\nu=\frac{4\pi^2 S \alpha\kappa^2}{M^2},
\label{eq:gdh}
\end{equation}
where $\sigma_{P}$ and $\sigma_{A}$ are helicity-dependent photoproduction cross sections, 
 $\nu$ is the probing photon energy, $\nu_0$ the pion production threshold, 
 $S$  the spin of the probed particle, $\kappa$ its anomalous magnetic moment and $\alpha$ is QED's coupling.
Like the Bjorken sum rule, the GDH sum rule was derived in a more general context that of QCD. However,
when the studied object is a nucleon or nucleus, QCD can be studied since the 
validity of Eq.~(\ref{eq:gdh}) depends on the high-$\nu$ behavior of $\sigma_{P}-\sigma_{A}$. 
Originally derived at $Q^2=0$, Eq.~(\ref{eq:gdh}) was later generalized to $Q^2>0$~\cite{Ji:1999mr}:
\vspace{-0.3cm}
\begin{equation}
\Gamma_1(Q^2)=\int_0^{x_0}g_1dx= \frac{Q^2}{2M^2} I_1(Q^2)
\label{eq:gdhsum_def2}
\vspace{-0.3cm}
\end{equation}
with $I_1$ the first polarized doubly virtual Compton scattering amplitude at $\nu \to 0$.

\begin{figure}[ht!]
\includegraphics[scale=0.37]{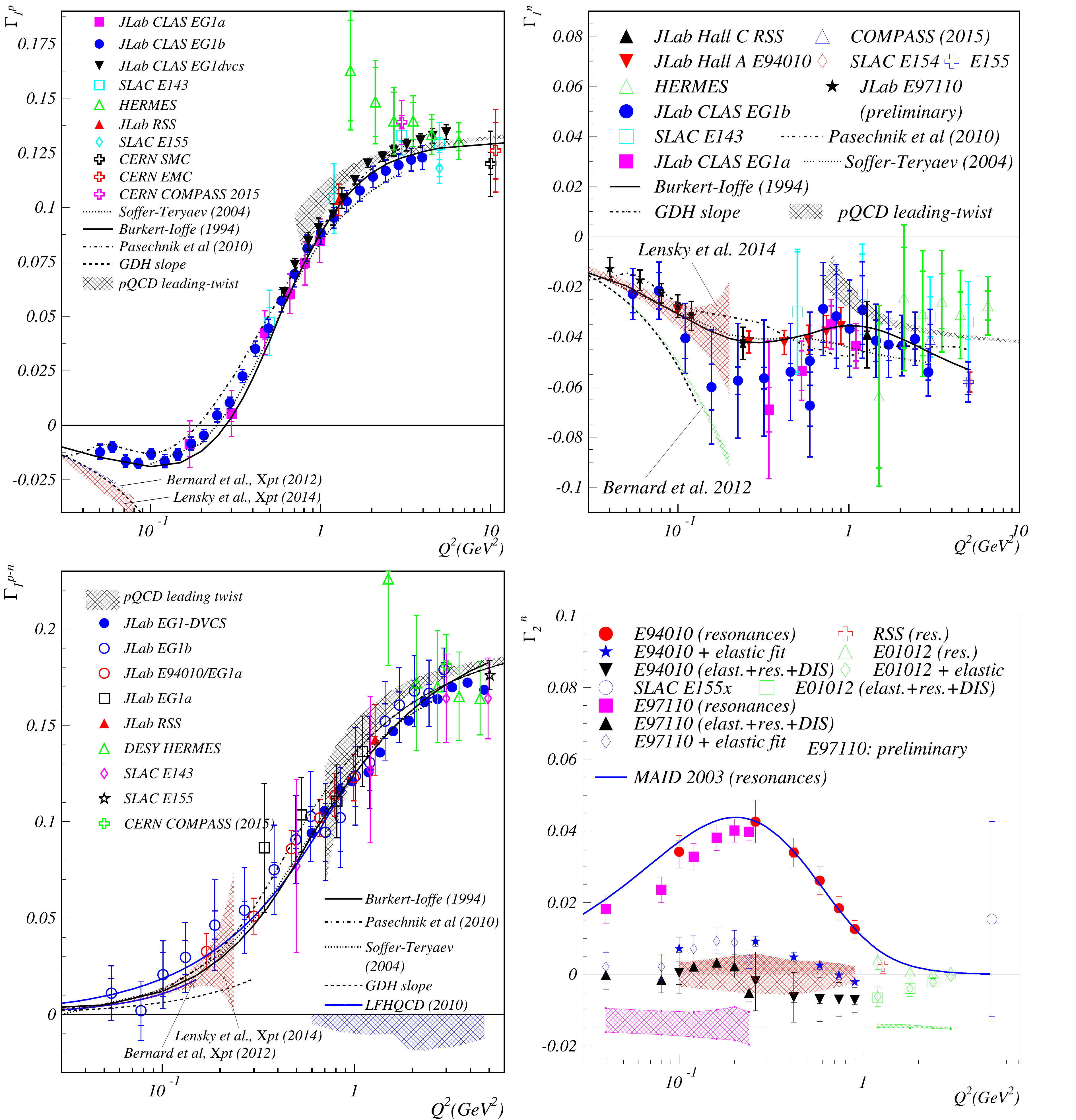}
\vspace{-0.3cm}
\caption{\small Moments $\Gamma_1^p$ (top left), $\Gamma_1^n$
(top right), $\Gamma_1^{p-n}$  (lower left) and $\Gamma_2^n$ (lower right).
%The gray band shows the leading-twist pQCD evolution. 
%$\chi$PT predictions are given by the continuous lines and bands at low $Q^2$. 
}
\label{fig:gamma1pn} 
\end{figure}

\begin{figure}[tb]
\includegraphics[width=7.6cm]{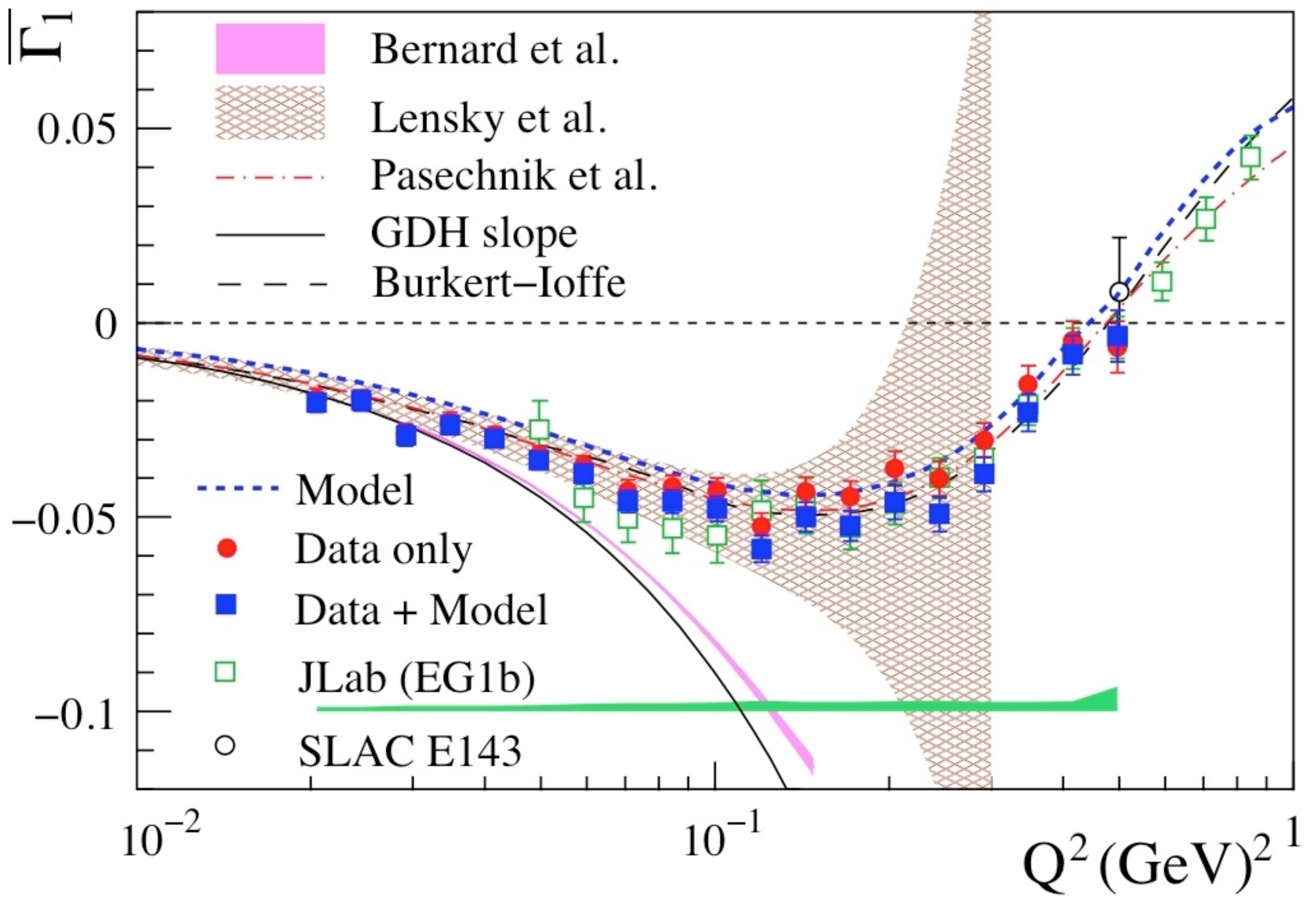}\includegraphics[width=6.3cm]{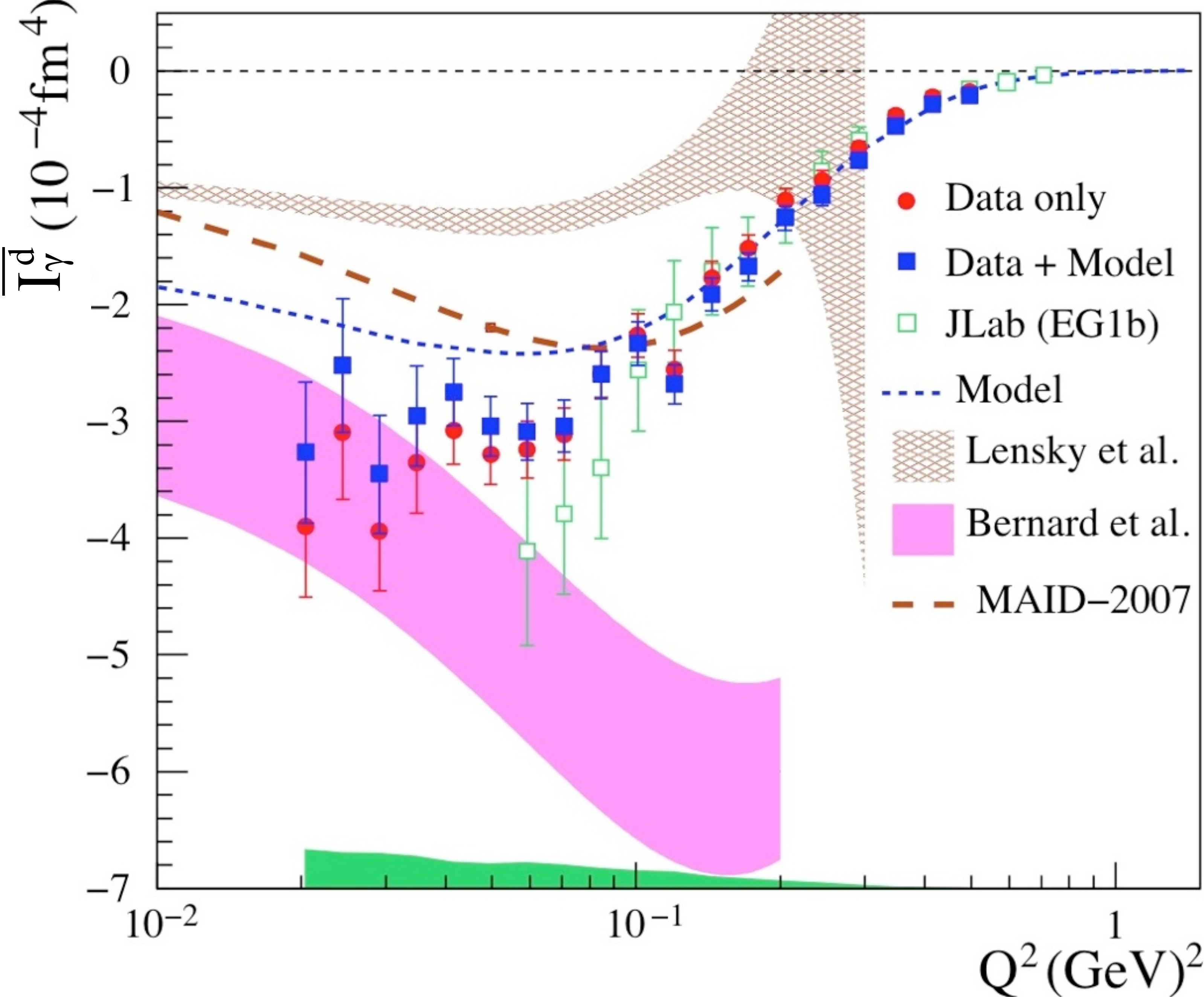}
\vspace{-0.5cm}
\caption{\small{Left: Deuteron's truncated moment $\bar{\Gamma}_1^d(Q^2)$ 
compared to models and $\chi$PT calculations. 
The error bars are statistical. The systematic uncertainty is provided by the horizontal band.
Right: truncated generalized spin polarizability $\bar{I}^d_\gamma(Q^2)$}. }
\label{Fig:EG4}
\end{figure}

Both the Bjorken and GDH sum rules involve $\Gamma_1$ and are thus two facets of a sum rule
holding at any $Q^2$. This offers a mean
to study the transition from hadronic to partonic d.o.f and to test nonperturbative approaches to QCD:
 $\Gamma_1$ can be measured at different $Q^2$ and then compared to the $I_1$ predicted by the nonperturbative approaches. 
 Data at $Q^2=0$ and large $Q^2$ come from BNL, MAMI, DESY, SLAC and CERN~\cite{Deur:2018roz}.
 The moderate $Q^2$ range was mapped at 
 JLab~\cite{Amarian:2002ar, Yun:2002td, Prok:2014ltt}. These data are shown on Fig.~\ref{fig:gamma1pn} 
 along with $\chi$PT predictions, the leading-twist pQCD evolution, and several
 models~\cite{Burkert:1992yk}-\cite{Drechsel:1998hk}.
A final $Q^2$  region remained to be mapped: that at low $Q^2$ 
where $\chi$PT predictions can be reliably tested, typically for $Q^2 \lesssim 0.1$ GeV$^2$. 
The lowest $Q^2$ points of the experiments~\cite{Amarian:2002ar, Yun:2002td}  did reach 
the $\chi$PT  region, but barely or with limited precision. 
The origin of the disagreement between some of the data and the corresponding $\chi$PT predictions 
was thus unclear. Hence, new experiments were run to measure  $g_1$, $g_2$ and their moments
well in the $\chi$PT  region: 
E97110 addressed the neutron,
E08-027 measured $g_2$ on the proton, and 
EG4 measured $g_1$ on the proton and deuteron~\cite{Adhikari:2017wox}.
The EG4 results on the deuteron are shown in Fig~\ref{Fig:EG4}. %Preliminary results 
%are available on the proton (EG4, E08-027) and neutron E97110.
%

\section{Bridging the hadronic and partonic regions \label{AdS/QCD}}

What does the $Q^2$ mapping of $\Gamma_1$  teach us about the connection between hadronic and partonic 
d.o.f? At high $Q^2$, higher twists can be extracted and related to confinement~\cite{Burkardt:2008ps},
the mechanism that makes hadronic d.o.f the relevant ones at low $Q^2$.
For example, the twist-4 coefficient $f_2^{p-n}$ was found to be relatively large~\cite{Deur:2004ti},
about half the leading twist contribution at $Q^2$ =1 GeV$^2$. This agrees with intuition:
confinement effects should be large at such $Q^2$. Twist-8 is also 
large but of opposite sign to that of $f_2^{p-n}$, suppressing the overall higher twist contribution
as the observation of parton-hadron duality requires. Other  higher twist studies
were done either by fit to moments, direct measurements of $g_2$ and $d_2$, or 
global analyses~\cite{Deur:2018roz}. Using Operator Product Expansion, higher twists 
allow us to apply pQCD to lower $Q^2$ than a leading twist analysis. 
At lower $Q^2$, $\chi$PT provides an effective theory for nonperturbative QCD. 
The agreement  between data and $\chi$PT predictions
\cite{Ji:1999mr,Ji:1999pd}-\cite{Lensky:2014dda} is ambivalent, 
depending on the observable and the $Q^2$ range on which the comparison is performed. Table 1
summarizes the state of affaires. The observables  $\delta_{LT}$ and $I^{p-n}_\gamma$ were 
expected to have robust $\chi$PT predictions because they have suppressed $\Delta_{1232}$ 
resonance contributions.
Furthermore, higher moments like $\delta_{LT}$, $I_\gamma$ or $d_2$ have little contribution from
the unmeasured low-$x$ region, and thus no low-$x$ extrapolation uncertainty. Nevertheless, 
$\chi$PT prediction failed, except for $\Gamma_1^{p-n}$. Recent calculations have however distinctly improved. 
\begin{table}[t]
\begin{center}
\footnotesize
\begin{tabular}{|c|c|c|c|c|c|c|c|c|c|c|}
\hline 
Ref. & $\Gamma_1^p$ & $\Gamma_1^n$ & $\boldsymbol{\Gamma_1^{p-n}}$ & $\Gamma_1^{p+n}$ &  $I^p_\gamma$ & $I^n_\gamma$ & $\boldsymbol{I^{p-n}_\gamma}$ & $I^{p+n}_\gamma$ & $\boldsymbol{\delta_{LT}^n}$ & $d_2^n$  \tabularnewline
\hline
\hline 
Ji 1999  \cite{Ji:1999pd,Ji:1999mr}  & X & X & A & X & - & - & - & - & - & - \tabularnewline
\hline 
Bernard 2002 \cite{Bernard:2002bs} & X & X & A & X & X & A & X & X & X & X\tabularnewline
\hline 
Kao 2002 \cite{Kao:2002cp}  & - & - & - & - & X & A & X & X & X &  X\tabularnewline
\hline 
Bernard 2012 \cite{Bernard:2012hb}  & X & X & A & X & X & A & X & X & X & -\tabularnewline
\hline 
Lensky 2014 \cite{Lensky:2014dda} & X & A &  A & A &  A & X & X & X &  $\sim$ A &  A\tabularnewline
\hline
\end{tabular}
\vspace{-0.2cm}
{\small
\caption{\small Comparison between data and $\chi$PT predictions. Observables for which $\chi$PT was 
expected to be most robust  are in bold.
 ``A" means that data and $\chi$PT agree up to 
$Q^2\approx0.1$ GeV$^2$, ``X" means a disagreement and ``-" that no prediction was available.
%The {\scriptsize \emph{p+n}} superscript indicates either proton+neutron data added together, 
%or data from deuteron without its break-up channel.
}}
\label{xpt-comp}
\end{center}
\end{table}

In all, the JLab nucleon spin structure data provided a precise mapping 
of the low and moderate $Q^2$ regions, triggering improvements in $\chi$PT predictions 
at low $Q^2$~\cite{Bernard:2012hb,Lensky:2014dda}, and in perturbative techniques 
at high $Q^2$~\cite{Pasechnik:2010fg}. In particular, $\alpha_s$ 
obtained from JLab's $\Gamma_1^{p-n}$ data~\cite{Deur:2005cf} motivated an AdS/QCD calculation of  
$\alpha_s$~\cite{Brodsky:2010ur} which lead to an analytical calculation of the hadron mass 
spectrum with $\Lambda_s$ as single input parameter~\cite{Deur:2014qfa}.  
The analytic calculation of the hadron masses from  $\Lambda_s$  is an ultimate goal of strong 
force studies. AdS/QCD remains a semi-classical 
approximation of QCD, but its calculation of masses  from $\Lambda_s$  is an 
encouraging step toward this goal. 

\section{Perspectives}
In the shorter term, the JLab nucleon spin structure program naturally expands 
to higher energy thanks to the recent 12 GeV upgrade of JLab. 
Precise  polarized PDF measurement will be extended to $x \approx0.8$.  
The data will be taken in 2019~\cite{large-x A_1 12 GeV exps}. In addition to the
new insight on $L_q$ given by these data, GPDs and TMDs will
constrain it more directly. GPDs and TMDs are a major aspect of the
12 GeV program of Halls A, B and C, with 11 GeV data already taken in Halls A and B. Furthermore, higher
$Q^2$ measurements of $g_1$ in Hall B will constraint $\Delta G$~\cite{Leader:2006xc} with different 
reactions than that of RHIC and thus be complementary to its $\Delta G$ program. 

In the longer term, the Hall A SoLID projet~\cite{SoLID} 
will yield GPDs and TMDs with high precision  
and over extended kinematics, while the EIC~\cite{Accardi:2012qut}  will  
measure polarized PDFs, GPDs and TMDs, and thus OAMs, at low $x$. 
Furthermore, its precise inclusive data measured over the extended $x$ reach will 
further constrain $\Delta G$. 

\section{Conclusion}
Nucleon spin structure study remains an active field of research. In the pQCD  region, 
JLab provides spin asymmetries at high-$x$,
$g_1$ and $g_2$, their various moments, and higher twists. The JLab  global
analysis JAM determines (un)polarized PDFs and higher twists. 
Explored at 6 GeV, GPDs and TMDs are a major aspect of the current 12 GeV program. 

At lower $Q^2$, JLab data complement the ones from CERN, SLAC and DESY
enabling the study of confinement and of the emergence of the effective hadronic degrees of freedom.
Most of the moderate $Q^2$ JLab data are now available, as are the low $Q^2$  deuteron data. 
The low $Q^2$ proton and neutron data are expected for 2019.

Arguably, JLab's spin sum rule program reached its goal: by providing a precise mapping at low and moderate $Q^2$, 
it triggered progress in theory, improving the description of the strong force from low $Q^2$ ($\chi$PT), 
through intermediate $Q^2$ (Schwinger--Dyson equations, AdS/QCD) 
and to high $Q^2$ (improved perturbative techniques). 
An example of such progress is  the analytic AdS/QCD calculation of the hadron masses from $\Lambda_s$.
The 12 GeV upgrade of Jlab allows for precision measurements of $g_1$ and $g_2$ at higher $x$ and $Q^2$, as 
well as higher twist, GPD and TMD extractions. Quark OAM and $\Delta G$ can then
be extracted from them.  In the longer term, the SoLID detector will improve GPD and TMD
measurements while the EIC will explore  the nucleon spin structure at very low $x$. 

%\Acknowledgements

\end{document}